# On-chip Real-time Hyperspectral Imager with Full CMOS Resolution Enabled by Massively Parallel Neural Network

Junren Wen[a,c,d], Haiqi Gao[a,c,d], Weiming Shi[a,d], Shuaibo Feng[a,d], Lingyun Hao[a,c,d], Yujie Liu[b], Liang Xu[b], Yuchuan Shao[a,c], Yueguang Zhang[b], Weidong Shen[b,*], and Chenying Yang[a,b,*]

[a] Hangzhou Institute for Advanced Study, University of Chinese Academy of Sciences, Hangzhou, Zhejiang, 310024, China

[b] State key laboratory of Modern Optical Instrumentation, Department of Optical Engineering, Zhejiang University, Hangzhou, Zhejiang, 310027, China

[c] Shanghai Institute of Optics and Fine Mechanics, Chinese Academy of Sciences, Shanghai, 201800, China

[d] Center of Materials Science and Optoelectronics Engineering, University of Chinese Academy of Sciences, Beijing, 100049, China

\* E-mail: adongszju@hotmail.com (Weidong Shen); ycheny@zju.edu.cn (Chenying Yang)
\* Telephone: +86 (571) 8795 1190 (Weidong Shen) ; +86 (571) 8608 7323 (Chenying Yang)

**Abstract**

Traditional spectral imaging methods are constrained by the time-consuming scanning process, limiting the application in dynamic scenarios. One-shot spectral imaging based on reconstruction has been a hot research topic recently and the primary challenges still lie in both efficient fabrication techniques suitable for mass production and the high-speed, high-accuracy reconstruction algorithm for real-time spectral imaging. In this study, we introduce an innovative on-chip real-time hyperspectral imager that leverages nanophotonic film spectral encoders and a Massively Parallel Network (MP-Net), featuring a 4 × 4 array of compact, all-dielectric film units for the micro-spectrometers. Each curved nanophotonic film unit uniquely modulates incident light across the underlying 3 × 3 CMOS image sensor (CIS) pixels, enabling a high spatial resolution equivalent to the full CMOS resolution. The implementation of MP-Net, specially designed to address variability in transmittance and manufacturing errors such as misalignment and non-uniformities in thin film deposition, can greatly increase the structural tolerance of the device and reduce the preparation requirement, further simplifying the manufacturing process. Tested in varied environments on both static and moving objects, the real-time hyperspectral imager demonstrates the robustness and high-fidelity spatial-spectral data capabilities across diverse scenarios. This on-chip hyperspectral imager represents a significant advancement in real-time, high-resolution spectral imaging, offering a versatile solution for applications ranging from environmental monitoring, remote sensing to consumer electronics.

## 1. Introduction

By capturing both spatial and spectral information, spectral imaging offers a comprehensive view that surpasses traditional digital imaging or point spectroscopy. This advanced technique finds extensive applications across various domains including environmental monitoring[1], precision agriculture[2–4], medical diagnosis[5], art restoration[6], remote sensing[7], and consumer technology. Conventional scanning methods (whiskbroom, pushbroom, and wavelength scan) provide exceptionally high spectral resolution alongside a broad wavelength range covering visible-near infrared region. However, the time-consuming scanning process limits the application in real-time or dynamic imaging scenarios. Addressing the speed constraints, exploration of snapshot spectral imaging technologies has become a growing focus among researchers. The strategies to achieve snapshot spectral imaging can primarily be divided into three categories: amplitude-coded methods (e.g. amplitude encoding with coded apertures[8,9]), phase-coded spectral imaging (e.g. phase encoding with diffractive optical elements[10,11]), and wavelength-coded spectral imaging (e.g. wavelength encoding with broadband spectral-filtering elements)[12]. Among these strategies, the scheme that combines broadband spectral encoders with AI-driven reconstruction algorithm presents superior spectral reconstruction accuracy and the potential for CMOS-compatible processing technology simultaneously. From an alternative perspective, wavelength-coded spectral imaging is essentially based on the concept of reconstructive spectrometers, which primarily involves tailoring the spectral responses, either directly at the detector level (e.g. nanowire[13,14], black phosphorus detector[15], perovskite detector[16], and van der Waals junction[17]) or through optical elements placed atop the detectors (e.g. quantum dot[18,19], metasurface[20], thin film[21,22], PC slab[23–25], and plasmonic tile[26]). To attain spectral imaging with high spatial resolution, modification of the optical elements over the detectors is more practical. By miniaturing the spectrometers to a micrometer scale and configuring them into arrays, one-shot spectral imaging becomes feasible.

Currently, the leading methods for spectral encoding employ metasurface arrays[27–30] and thin-film based Fabry-Pérot (F-P) cavities[31], with two main approaches to positioning the spectral encoders in relation to the CMOS image sensor (CIS) pixels: one-to-one mapping where each encoder is matched with a single pixel, or one-to-many correspondence. The one-to-one mapping demands extremely high alignment precision and notably small size for each spectral encoder (dimension smaller than 5 μm × 5 μm is required typically). Conversely, the one-to-many correspondence simplifies requirements for alignment and encoder size, but results in a compromise on spatial resolution. Moreover, for more precise spectral reconstruction, it is imperative to increase the quantity of spectral encoders within each individualized micro-spectrometer, while simultaneously ensuring the low correlation of the spectral responses. This requirement introduces an inherent trade-off between the complexity of fabrication and the accuracy of spectral reconstruction. Considering the algorithmic aspect, utilizing AI-driven Spectral Encoder and Decoder (SED) reconstruction network for one-shot spectral imaging applications[28,32–34], the weight matrix for spectral encoding is generally fixed and corresponds to the average transmission spectra of the spectral encoders for all micro-spectrometers. However, the configuration struggles to adequately compensate for the discrepancies in transmittance among the micro-spectrometers.

Addressing the aforementioned tackles, we propose an on-chip real-time hyperspectral imager with full CMOS resolution based on nanophotonic film spectral encoders alongside a highly parallelized reconstruction algorithm. Each individualized micro-spectrometer comprises a 4 × 4 nanophotonic compact all-dielectric film units, with each unit covering 3 × 3 CIS pixels. The curved film structure for each unit allows for varied modulation of incident light across the underlying CIS pixels, leading to unique encoding for the overall 12 × 12 pixels. Increasing the number of spectral encoders for single micro-spectrometer from 16 to 144 can significantly enhance the accuracy of spectral reconstruction. The fabrication of the micrometer-scale nanophotonic film array is characterized by its cost-effectiveness, repeatability, and capability for mass, large-area production. Moreover, we introduce Massively Parallel Network (MP-Net), an enhanced SED that specifically address the variability in transmittance, thereby enhancing the accuracy of imaging. Hyperspectral imaging experiments in both indoor and outdoor environments for static and moving objects are performed, validating the prominent hyperspectral imaging capabilities of the proposed imager.

**2. Results**

2.1 Hardware of the hyperspectral imager

The hardware of the hyperspectral imager consists of the micrometer-scale nanophotonic film array and the CMOS image sensor (CIS) beneath, as shown in Fig.1 (a). Specifically, the film array spans about 1850 × 2150 pixels of the CIS, encompassing approximately 67% of the total area. The single individualized micro-spectrometer is composed of a 4 × 4 nanophotonic compact all-dielectric film units, with distinctive spectral responses arising from variations in the film's morphology, shown in Fig.1 (b-c). Each unit has a size of 7.2 μm × 7.2 μm covering 3 × 3 CIS pixels. The microscope image of the nanophotonic filter array is shown in Fig. S1, Supporting Information. In this study, we set only one thickness-variant layer to facilitate efficient spectral encoding and to enhance manufacturing feasibility, as shown in Fig.1 (c). The bottom and top three layers are fabricated by electron beam evaporation, whereas the intermediate variable layer is fabricated through a process involving four lift-off cycles with pattern transfer technique of the film deposition. The detailed description of the fabrication cycle is shown in Fig. S2, Supporting Information. By modifying the mask pattern as well as the deposition time, we produce divergent thicknesses for the 16 nanophotonic film units, as detailed in Fig. S3, Supporting Information. Due to the shadow effect during the thin film deposition process, a curved nanophotonic film structure can be obtained for each unit. Details of the structure, as characterized by Atomic Force Microscope (AFM) and Scanning Electron Microscopy (SEM), are presented in Fig. S4 and Fig. S5, Supporting Information, respectively. Such structural configuration enables different incident light modulation among the underlying 3 × 3 CIS pixels. Additionally, by carefully adjusting the morphology of the photoresist (by finely tuning the exposure dose), the spectral variation among the underlying pixels is increased. As a result, the number of effective spectral encoders for a single micro-spectrometer increases to ninefold, transitioning from a 4 × 4 array (with the dimension of each unit 7.2 μm × 7.2 μm) to a 12 × 12 array (with the dimension of each unit 2.4 μm × 2.4 μm), significantly enhancing the spatial resolution to the level of CMOS.

Fig.1 (d) presents the transmission spectra heatmap for the 16 nanophotonic film units measured by a commercial spectrophotometer. The dimension of the testing light spot is approximately 7 μm × 7 μm, indicating that the transmission spectra represent an averaged spectral response from the 9 sub-spectral encoders. Part of the transmission spectra are intuitively illustrated in Fig.1 (e), and notably, these responses are characterized by multi-peak-valley and broadband encoding features. The transmission spectra of the 16 nanophotonic film units recorded at various locations within the 5 mm × 5.5 mm filter array are shown in Fig. S6, Supporting Information. Spectral randomness of the 301 spectral channels ranging from 400nm to 700nm with 1nm resolution, and correlation coefficients of the 16 nanophotonic film units are shown in Fig.1 (f) and Fig. S7, Supporting Information, respectively. Although there is a general increase in correlation, the number of the lift-off cycles has been reduced to one-fourth compared to the analogous scheme reported in Reference[21].

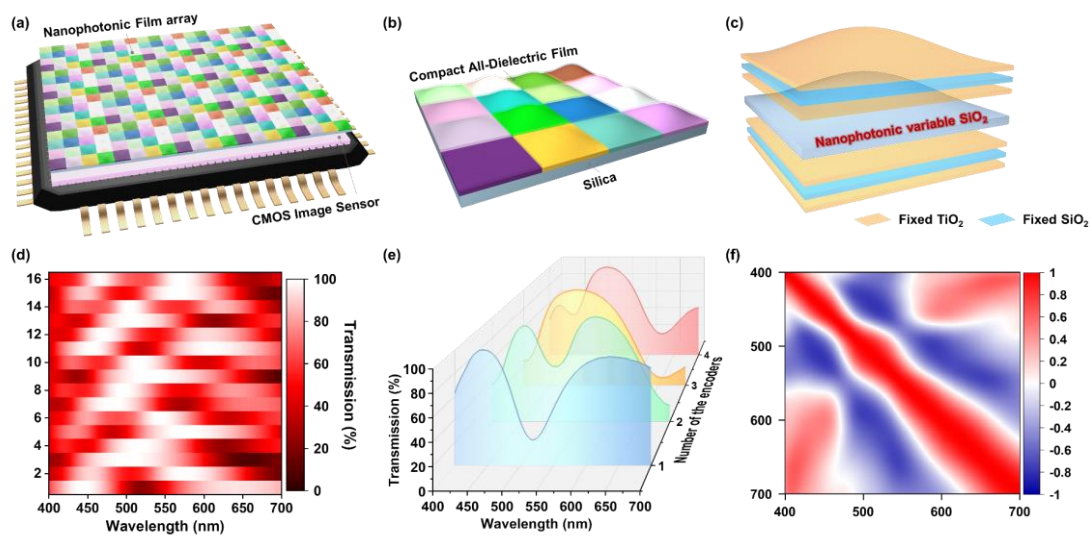

Fig. 1 Full-CMOS-resolution on-chip hyperspectral imager. (a) Schematic diagram of the hyperspectral imager, featuring a micrometer-scale nanophotonic film array positioned above the CMOS image sensor (CIS). (b) Schematic diagram of the 4 × 4 nanophotonic film array. The compact all-dielectric film is deposited of on silica substrate with a thickness of 500 μm. (c)

Schematic diagram of the single nanophotonic 7-layer film structure. High and low refractive index materials ($TiO_2$ and $SiO_2$) are stacking alternatively. The curved intermediate layer is fabricated through four lift-off cycles, with each cycle consisting of patterning, deposition, and stripping. (d) The transmission spectra heatmap for the 16 nanophotonic film units. (e) Part of the transmission spectra curves of the units, characterized by multi-peak-valley and broadband encoding features. (f) Spectral randomness of the 301 spectral channels ranging from 400nm to 700nm with 1nm resolution for the 16 nanophotonic film units.

## 2.2 Real-time Hyperspectral Imaging with MP-Net

In previous work, the AI-driven Spectral Encoding and Decoding (SED) reconstruction network is widely applied for computational spectroscopy and spectral imaging. However, when the algorithm is applied for one-shot spectral imaging application (with millions of micro-spectrometers in total), the fixed weight matrix for spectral encoding, which serves as a proxy for the spectral responses of the encoders, varies among different micro-spectrometers. Assigning the weight matrix specifically tailored to a particular micro-spectrometer, will lead to reconstruction errors as demonstrated in Fig. S8, Supporting Information. The discrepancies of the intensities across different micro-spectrometers arise partly due to the variations intrinsic to the fabrication process, and partly due to misalignment, as detailed in Fig. S9, Supporting Information.

To address this issue, assigning each micro-spectrometer a unique spectral reconstruction network, tailored to its specific spectral responses, the reconstruction fidelity and imaging performance will be significantly enhanced. However, in our work, approximately 600 × 700 = 420, 000 individualized micro-spectrometers exist, and the total number will be dramatically augmented when considering multiplexing of the adjacent pixels. Sequentially training the millions neural networks is a time-consuming task, potentially extending over several weeks. Therefore, we comprehensively considered the integration of the aforementioned two approaches. Employing specific neural network for multiple micro-spectrometers within a constrained area, and additionally, based on the distribution of the entire spectrometers, we establish distinct regions, enabling the deployment of massively parallel network for parallel training and reconstruction.

Workflow of the hyperspectral imaging system is shown in Fig. 2(a), and the illuminated object is spatially and spectrally encoded by the on-chip hyperspectral imager. The encoding procedure can be expressed as

$$I_i(x,y) = \int_{\lambda_{min}}^{\lambda_{max}} I(x,y,\lambda)D(\lambda)S(x,y,\lambda)T_i(\lambda)d\lambda, \ i = 1,2,\ldots,144$$

where $I(x,y,\lambda)$ refers to the normalized spectrum of the illumination, $D(\lambda)$ refers to the normalized spectral response of the CIS, $S(x,y,\lambda)$ represents the spectrum under test and $T_i(\lambda)$ represents the spectral responses of the 144 spectral encoders. $I_i$ denotes the intensity vector for single micro-spectrometer, and all $I_i$ values are embedded within the raw image.

We segment the raw image into a total of 14 × 16 sub-regions (a portion of the peripheral pixels has been discarded), with each encompassing an area of 132 × 132 CIS pixels, shown in Fig. 2(b) and Fig. S10, Supporting Information. In order to achieve imaging resolution that reaches the level of the CIS, the spectral encoders of the adjacent micro-spectrometers are overlapped. Within each sub-region, a single weight matrix (corresponding to the spectral responses of the micro-spectrometer situated at the central point) is applied to all multiplexed micro-spectrometers. Leveraging the segmentation of the raw image by MP-Net, rigorous and precise alignment between the nanophotonic array and the CIS is not needed, allowing for the straightforward manual placement.

Layers 1-3 function as hidden layers and Layer 4 serves as the output layer in the proposed Massively Parallel Network (MP-Net). These layers are segmented in alignment with the sub-regions of the raw image. Within each sub-region across these layers, as indicated by the same color in Fig. 2(a), the neurons are interconnected exclusively within the respective sub-region and do not establish connections with others. This design effectively prevents crosstalk between data from different sub-regions, ensuring that the parameters of each sub-neural network are specifically tailored and optimized by the data within the particular area. The detailed description of MP-Net is shown Fig. 2(c), where N and n denotes the 14 × 16 sub-regions and the number of micro-spectrometers in a sub-region, respectively. In Layer 1, each sub-neural network, corresponding to each sub-region, is capable of receiving multiple parallel inputs (akin to batch processing), with each input comprising a vector of dimensions (1, 144). Comparison of the MP-Nets, differentiated by each sub-neural network in Layer 1 with 16 or 144 input neurons — representative of the number of spectral encoders in each micro-spectrometer — is shown in Fig. S11, Supporting Information. The entire architecture of the MP-Net is "(N×144)-BN-LR-FC-(N×500)-BN-LR-FC-(N×500)-BN-LR-FC-(N×301)-BN-LR". FC represents a fully connected layer, BN represents a Batch Normalization layer, and LR denotes a Leaky Rectified Linear Unit, while numbers indicate the number of neurons for each layer. It is worth noting that within the MP-Net, both FC and BN units have been restructured to accommodate a three-dimensional connection. The redefinition extends the dimensions from a 2D connection of 144 × 500 to a 3D one of N × 144 × 500 (taking the first FC as an example), allowing the network to process N parallel inputs. Such architectural adaptation is crucial for handling the complex multi-dimensional data within the network, thereby increasing the efficiency and performance of the hyperspectral imaging system. Finally, the reconstructed data is reshaped into a datacube which has a spatial resolution equivalent to that of the CMOS.

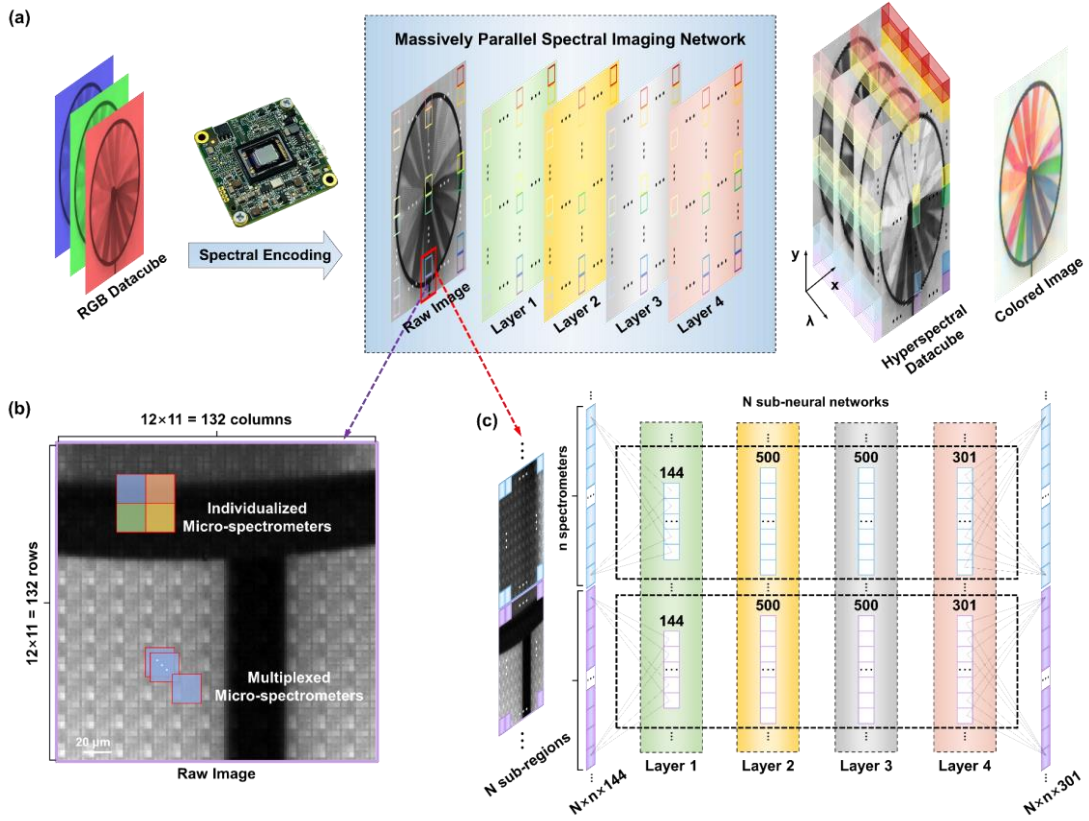

Fig. 2 Massively Parallel Network (MP-Net) for high-precision hyperspectral imaging. (a) Workflow of the hyperspectral imaging system. (b) A sub-region in the raw image. It covers an area of 132 × 132 CIS pixels, and contains 11 × 11 individualized micro-spectrometers. By multiplexing the adjacent pixels, the number of micro-spectrometers significantly increases, and the spatial resolution for hyperspectral imaging reaches the level of CMOS. (c) Schematic of the MP-Net. The entire architecture is "(N×144)-BN-LR-FC-(N×500)-BN-LR-FC-(N×500)-BN-LR-FC-(N×301)-BN-LR", where N and n denote the number of the sub-regions (sub-neural networks) and the number of the multiplexed micro-spectrometers, respectively. In our work, N = 14 × 16 = 224 and n = 132 × 132 = 17424. Such sequence represents the layers and processes within the network: Batch Normalization (BN), Leaky Rectified Linear Unit (LR), and Fully Connected (FC) layers. Both FC and BN units are restructured to extend the dimensions from a 2D connection of 144 × 500 to a 3D one of N × 144 × 500 (taking the first FC as an example).

To evaluate the performance of MP-Net, numerical simulations were performed using the hyperspectral image (HSI) dataset CAVE[35] and ICVL[36]. The original dataset is encoded by the experimentally obtained spectral responses of the on-chip spectral imager, and the MP-Net is trained via merely HSI dataset CAVE or ICVL, ensuring that there is no overlap between the training and testing datasets. The noise level is set as 0 in the simulation procedure, and the focus remains on assessing the inherent capability of MP-Net to reconstruct images accurately, without the confounding effects of additional noise. The ground truth and the reconstructed images visualized in RGB form are shown in Fig. 3(a) and Fig. 3(b), respectively. To evaluate the reconstruction quality and accuracy, key metrics including Peak Signal-to-Noise Ratio (PSNR), Structural Similarity Index Measure (SSIM), and the average Mean Squared Error (MSE) between the reference and the reconstructed images are calculated. For every HSI image in CAVE and ICVL,

PSNR exceeds 35dB and SSIM consistently surpasses 0.95, indicating a high level of fidelity in the spectral imaging process. Furthermore, the average MSE is also found to be within the $10^{-4}$ magnitude, demonstrating remarkably minimal deviation from the reference spectra. To visually depict these minor discrepancies, an error map is provided in Fig. 3(c), and we can find that the reconstruction error originates from the areas with higher brightness levels. Moreover, we randomly select three points (with the RGB patches in Fig. 3(a)) for each HSI and the MSEs between the reference (the black dashed lines) and the reconstructed spectra (the red solid lines) range from $1.49\times10^{-5}$ to $4.87\times10^{-4}$ with an average value of $1.40\times10^{-4}$, shown in Fig. 3(d). Based on the results, by providing the MP-Net with a sufficiently general training dataset, high-precision imaging of common objects and scenes can be achieved.

Additionally, the performance of MP-Net is assessed in terms the frame rate of spectral imaging. For a hyperspectral datacube with the dimension of 512 × 512 × 301, the reconstruction time is recorded at 0.33 seconds with single commercially available graphics processing unit (GPU: Nvidia GeForce RTX4090). Furthermore, when recovering a datacube with fewer spectral channels (the dimension of the datacube is 512 × 512 × 31), the time significantly decreases to ~ 0.03 seconds, which is comparable with the processing time in the previous work[31]. This rate exceeds 30 frames per second (fps), which aligns with the maximum frame rate of the adopted CIS, showcasing the capability for real-time and video-rate spectral imaging.

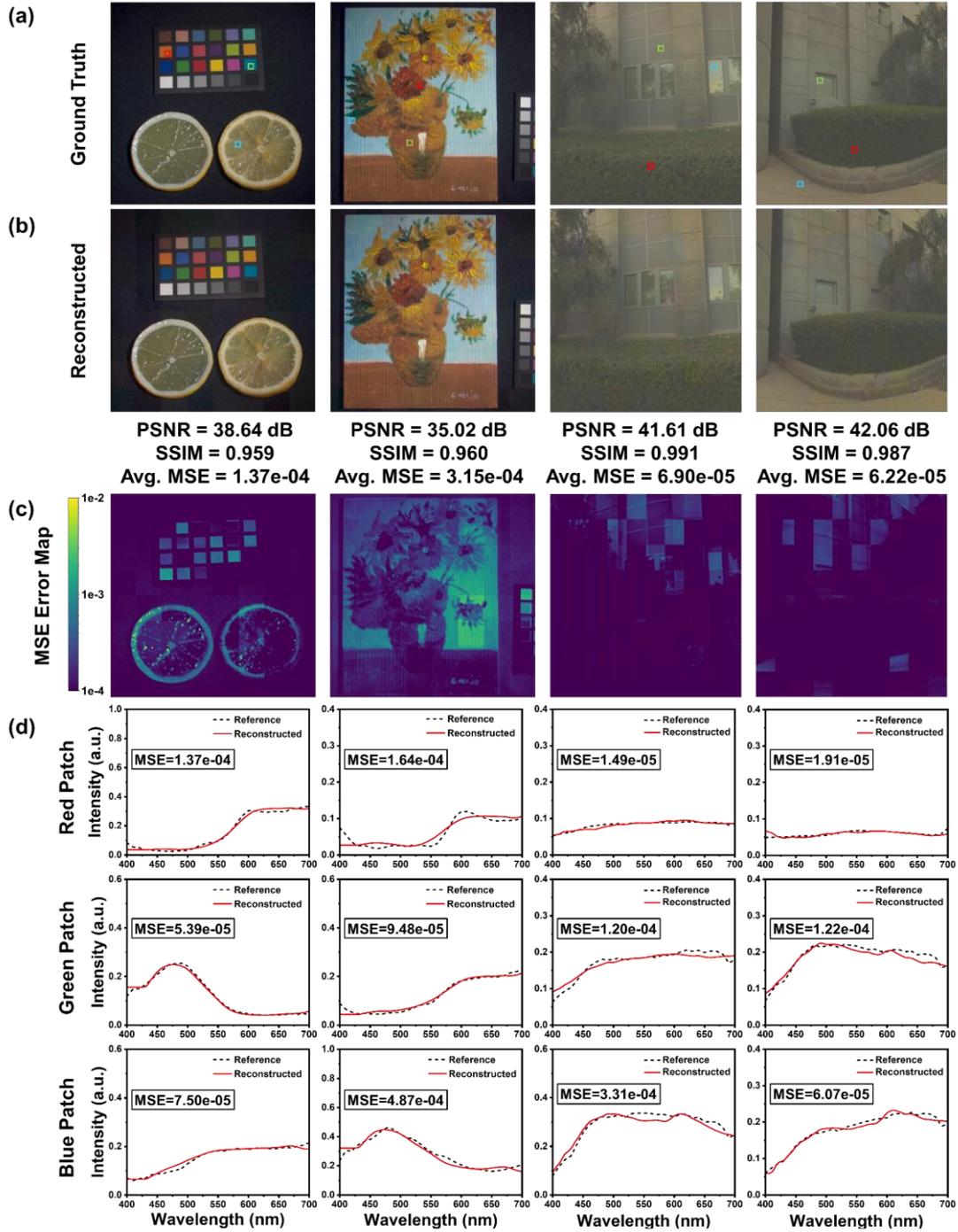

Fig. 3 Hyperspectral images (HSIs) reconstruction using MP-Net. (a) The ground truth images in the HSI dataset CAVE and ICVL. (b) The reconstructed hyperspectral images visualized in RGB form. The reconstructed images exhibit no significant differences compared to the reference ones. Below the images, PSNR, SSIM and the average MSE are calculated for each HSI image. (c) The error map showcasing the MSE of each pixel in the HSI. The color bar is set in a logarithmic scale, ranging from $10^{-2}$ to $10^{-4}$. (d) The reference (the black dashed lines) and the reconstructed spectra (the red solid lines) of the randomly selected plots denoted in Fig. 3(a). The value in the figure indicates the MSE between the two spectra.

2.3 Hyperspectral imaging of static and moving objects in both indoor and outdoor environments

We conduct experimental evaluations of the on-chip hyperspectral imager in both indoor and outdoor environments to thoroughly assess its performance under varying lighting conditions. Indoor illuminations are typically controlled, artificial lighting, while outdoor environments are characterized by natural and changing conditions. By assessing the capabilities in these distinct scenarios, we aim to demonstrate its adaptability and robustness in accurately capturing spectral data across a diverse range of illuminations, thus underscore the versatility and reliability of the on-chip hyperspectral imager in different real-world applications.

The indoor one-shot spectral imaging results are shown in Fig. 4. The imager operates in the active mode, where the absolute reflection of each image point is reconstructed regardless of the illumination. The detailed procedure for calibration is shown in Note 1, Supporting Information. Noises in spectral imaging predominantly arise from the detectors and the image acquisition process. To address this, we intentionally introduce the noise factor in training MP-Net to enhance the overall robustness and fidelity of spectral imager. A predetermined noise level is set, and noise, proportional to the calculated intensities and scaled by this level, is generated from a standard normal distribution $N(0, 1)$ and subsequently added to the original intensities. In Fig. 4 (a), the ground truth and the reconstructed images visualized in synthetic RGB format of the 24-color chart (Calibrite, Color Checker Classic Mini) are presented. We varied the noise levels from 0 to 0.2 to identify the optimal noise level for spectral imaging. The reference and the reconstructed spectra at different noise levels for the red, green and blue (RGB) patches are shown in Fig. 4 (b). The colored solid lines are the average reconstruction spectra of $150 \times 150$ image points in the particular color patch. We assume that the color chart is uniform and replicate the reference spectrum to match the size of $150 \times 150$. Therefore, the average MSE, PSNR and SSIM are calculated for the RGB patches, shown in Fig. 4 (c). We find that when the noise level is set as 0.1, the reconstructed images achieve the best quality, considering a comprehensive evaluation of the aforementioned metrics. This preferred noise level is applied in subsequent experiments for improved spectral imaging results.

Moreover, spectral imaging of the Negative 1951 USAF Test Target (Thorlabs R3L3S1N) is also performed indoor to evaluate the spatial resolution, shown in Fig. 4 (d). The 76.2 mm × 76.2 mm target covers a region of approximately $1450 \times 1700$ pixels, and the reconstructed image is shown in the leftmost figure. The enlarged reconstructed image (within the red dashed square) as well as the raw image captured by the hyperspectral imager of group -1 are shown in the middle, and the single line in element 6 occupies about 15 pixels on the raw image, which is comparable to the side length of single micro-spectrometer (12 pixels). This result indicates that the highest resolution limit has been attained under the current dimension of the micro-spectrometer. By switching the lens of the imager with a shorter focal length of 8 mm, the field of view (FOV) decreases, resulting in enlarged objects (within the blue dashed square) on the sensor, shown in the rightmost figure. In this case, element 3 of group 2 can be distinguished, reaching the spatial resolution of 5.04 lines per millimeter (lp/mm).

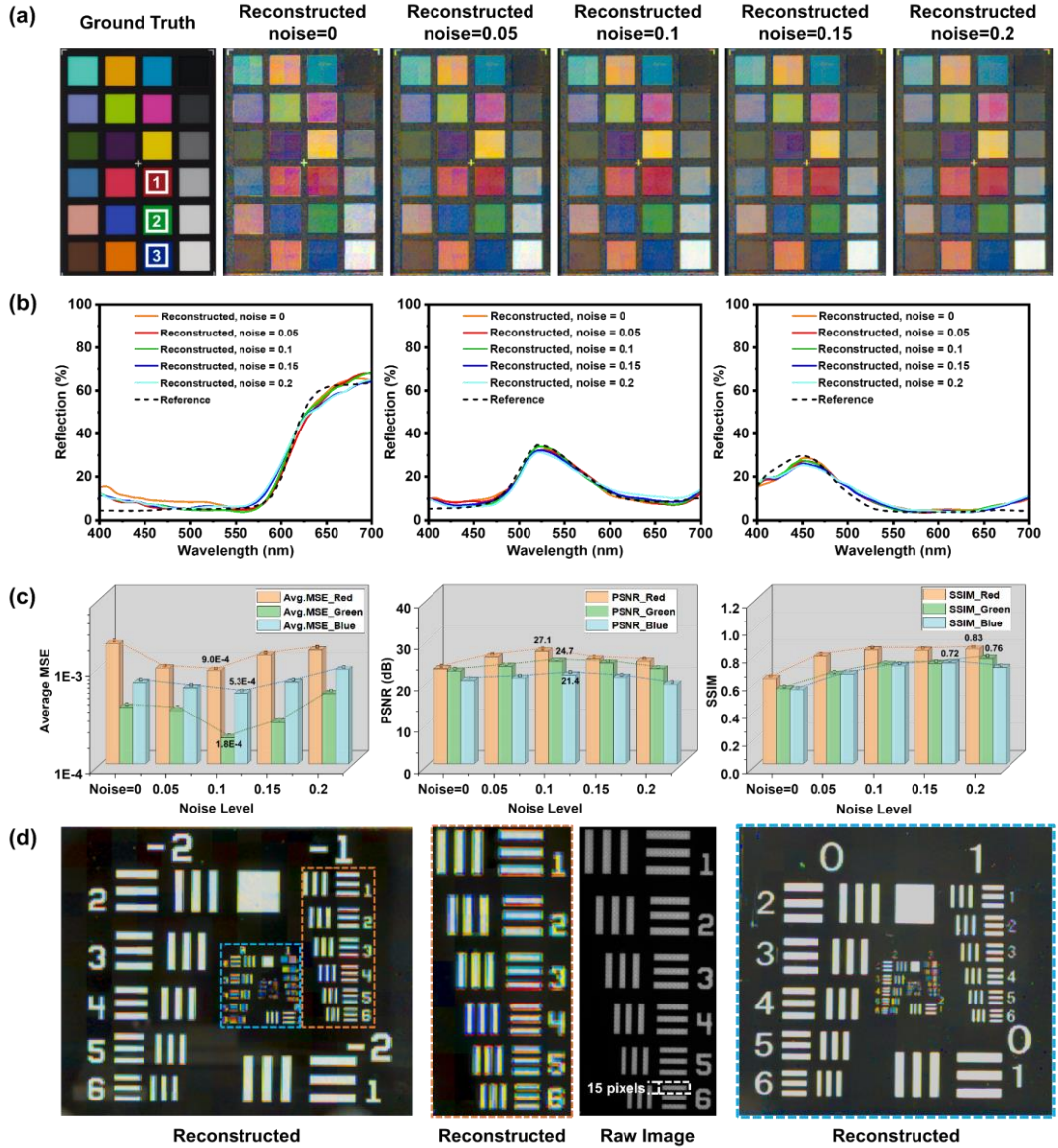

Fig. 4 Indoor one-shot hyperspectral imaging. (a) The ground truth and the reconstructed images visualized in synthetic RGB format of the 24-color chart. The predetermined noise levels of the reconstructed images are 0, 0.05, 0.1, 0.15, and 0.2, respectively. (b) The reference (the black dashed lines) and the reconstructed spectra (the colored solid lines) of different noise levels for the red, green and blue (RGB) patches. The reconstructed spectra are the average spectra of 22,500 (150 × 150) image points in the particular color patch. (c) Calculated average MSE, PSNR and SSIM for the RGB patches. Setting the noise level at 0.1 enables the reconstructed images to achieve optimal quality. (d) Spectral imaging of the Negative 1951 USAF Test Target. The reconstructed image is shown in the leftmost figure. The enlarged reconstructed image (in the red dashed square) and the raw image of the same area is shown the middle figure. The maximum spatial resolution reaches 5.04 lp/mm when the FOV focus on the central region in the blue dashed square, shown in the rightmost figure.

In outdoor environments with natural sunlight, we conduct hyperspectral imaging on a variety of subjects, including both static scenes and moving objects, shown in Fig. 5. Specifically, Fig. 5(a) illustrates hyperspectral imaging of a scene, which is composed of colored pencils and a tangram. The ground truth is captured by an RGB camera with the identical pixel numbers of the one used in the hyperspectral imager. The MP-Net is trained with noise level of 0.1. Color difference between the two images is observed, a phenomenon that can be ascribed to the illumination information included within the reference image. The reference and average reconstructed spectra (10 × 10 image points) of the Red/Yellow/Green patches in the tangram are also illustrated in the rightmost figure, where the MSEs are calculated and shown in the plot legend. The reconstruction MSEs for the three patches are $3.9\times10^{-3}$/ $6.0\times10^{-3}$/ $2.7\times10^{-3}$, respectively, and the values are slightly higher compared to those obtained in indoor environments. This distinction in MSEs can be attributed to the more variable and complex illumination conditions for outdoor environments, which introduce additional challenges in accurately reconstructing the absolute spectral characteristics. However, the relatively low MSEs demonstrate the robustness and adaptability of the reconstruction process in outdoor scenarios, affirming its efficacy in diverse illumination conditions.

Furthermore, to demonstrate the real-time imaging capability of the proposed hyperspectral imager, we present hyperspectral imaging of a rotating windmill, shown in Fig. 5(b). The raw images are recorded at 20 fps, and the maximum imaging frame rate can be increased to 30 fps which is the upper limit of the selected CIS. Reconstructed images in synthetic RGB form at six time points (0.0 s, 0.6 s, 1.2 s, 1.8 s, 2.4 s and 3.0 s) are shown in the leftmost figure, and the average rotation speed for the windmill is approximately 2 rad/s (~20 r/min). The middle and the rightmost figures depict the reconstructed monochromatic images of the rotating windmill at 450 nm and 550 nm. The dimension of the reconstructed hyperspectral datacube is 1210 × 1210 × 301 (about 1.46 million pixels), while the imaging time is approximately 0.59s for all spectral channels and 0.0020 s for a single channel. By downsampling the results to 21 spectral channels (15 nm each), the imaging frame rate reaches 24.4 fps, which surpasses the recorded frame rate.

Also, hyperspectral imaging of the flat moving 24-color chart at the speed of approximately 6 cm/s are shown in Fig. 5(c). We compare the reconstructed spectra of the RGB patches during the imaging process, shown in the rightmost image. The colored solid lines depict the average reconstructed spectra over multiple time points, while the shaded areas encircling the solid lines reveal minimal fluctuations of the reconstruction spectra. The result effectively demonstrates the consistency of the reconstructed spectra among the real-time spectral imaging procedure, highlighting the robustness of MP-Net and the capacity to maintain reconstruction accuracy of the proposed hyperspectral imager despite the potential temporal and spatial variations.

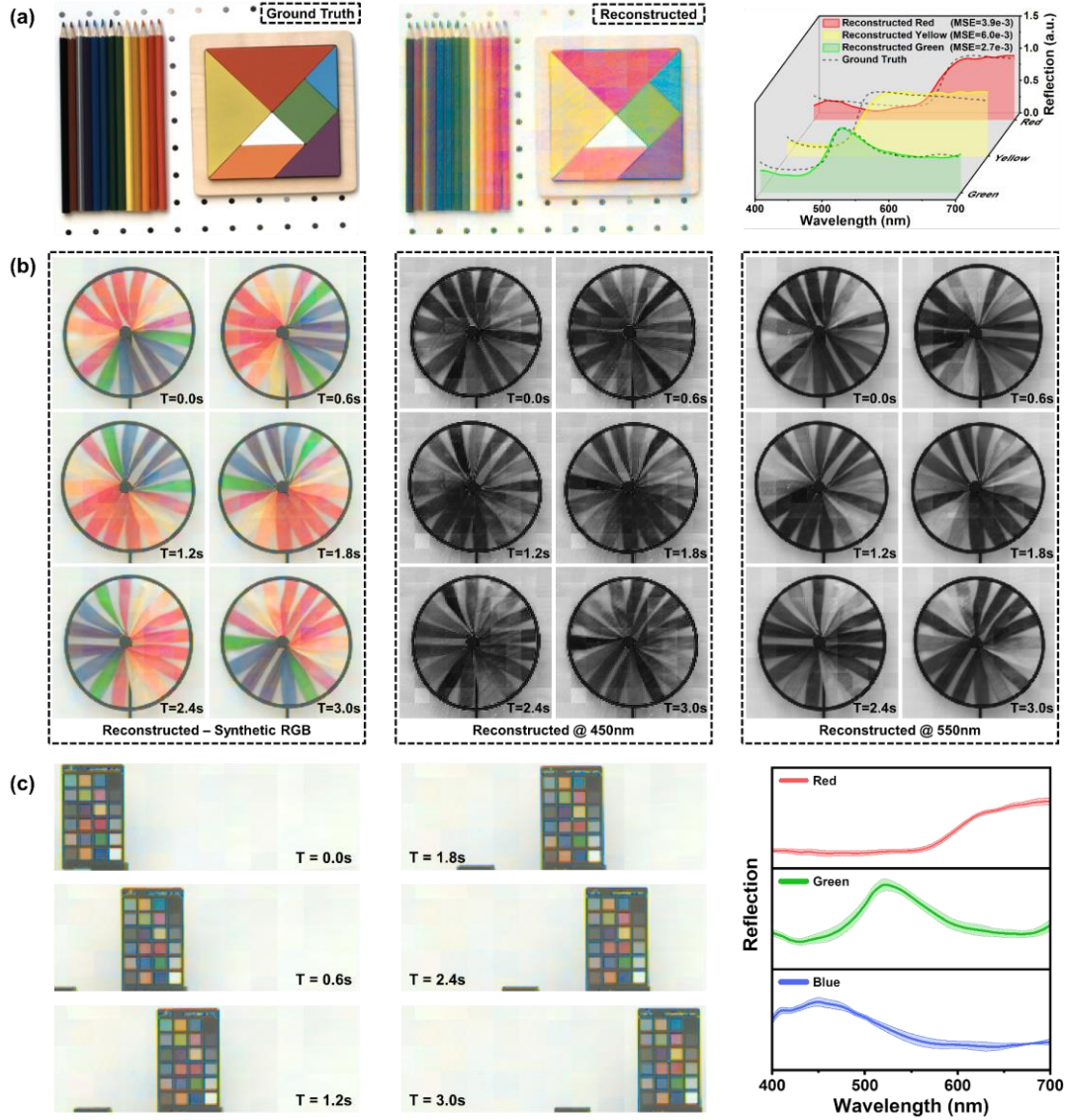

Fig. 5 Outdoor real-time hyperspectral imaging of static scenes and moving objects. (a) Snap-shot hyperspectral imaging of an outdoor scene including colored pencils and a tangram. The ground truth and the reconstructed images are shown in the leftmost and middle figures, respectively. The rightmost figure depicts the reference (the black dashed lines) and the reconstructed (the colored solid lines) reflection of the denoted RGB patches, and MSEs are calculated and shown in the plot legend. (b) Hyperspectral imaging results of a rotating windmill at the speed of approximately 2 rad/s (~20r/min), with the raw images captured at 20 fps. The imaging frame rate can reach 24.4 fps for reconstructing the 1210 × 1210 × 21 datacube with 21 wavelength channels. (c) Hyperspectral imaging results of a flat moving color board at the speed of approximately 6 cm/s, shown in the leftmost and the middle figures. The reconstructed spectra of the RGB patches within the imaging procedure are shown in the rightmost image. The colored solid lines depict the average reconstructed spectra over the imaging process, while the shaded areas encircling the solid lines reveal minimal fluctuations of the reconstruction spectra.

## 3. Discussion

In summary, we propose an on-chip real-time hyperspectral imager enabled by micrometer-scale nanophotonic film encoders and massively parallel network (MP-Net). The single individualized micro-spectrometer is composed of a 4 × 4 compact all-dielectric film units (each encompassing 3 × 3 CIS pixels). The formation of the curved nanophotonic film unit is jointly contributed by the shadow effect during the film deposition process, the engineered control of the exposure for UV lithography, and the misalignment between the filter array and the CIS. The curved nanophotonic structure facilitates the modulation of divergent incident light among the underlying CIS pixels. Therefore, a unique one-to-one mapping between the spectral encoders and the CIS pixels has been achieved by means of MP-Net configuration, enhancing the spatial resolution to that of the CMOS. To further reduce the impact of the incident angle on the spectral encoding, spectral encoders which are insensitive to the incident angle can be adopted[37,38]. During the training process of MP-Net, additional noise is incorporated to enhance the robustness of the reconstruction network. Hyperspectral imaging of static and moving objects in both indoor and outdoor environments are performed, demonstrating the robust adaptability of the imager and proficiency to acquire high-fidelity spatial-spectral datacube across a diverse range of scenarios.

Though minor manufacturing errors, e.g. misalignment for the photolithographic overlay process, shadow effect and non-uniformity for the thin film deposition, lead to discrepancies in the spectral responses of different micro-spectrometers among the filter array, the proposed MP-Net effectively addresses and eliminates these inconsistencies, thus enhancing the imaging quality. Moreover, MP-Net significantly reduces the necessity for rigorous and accurate alignment between the nanophotonic film array and the CIS by segmenting the raw image into sub-regions and providing specialized sub-neural network for each sub-region. Nevertheless, we find that block-like artifacts appeared across different sub-regions, a phenomenon attributable to the distinct weight matrices used for training of the sub-neural networks. Addressing the defect, we assign identical initial hyperparameters to the differently weighted sub-neural networks, and also employ transfer learning for training MP-Net. Unfortunately, these strategies does not result in significant improvements. Future research can explore alternative approaches to mitigate the block-like artifacts, e.g. advanced regularization techniques during the training procedure. Moreover, various network structures such as Transformer and Retentive Network (RetNet) can be applied for exquisite hyperspectral imaging performance.

## 4. Materials and methods

4.1 Design of the nanophotonic spectral encoders

The spectral responses of the nanophotonic spectral encoders are predesigned to verify the reconstruction performance using MP-Net. The spectral response generator based on the thin film transfer matrix method is implemented in Matlab, and the reliability of the generator is validated by a commercial software, e.g. optiLayer. The refractive index (n) and the extinction coefficient (k) employed in the simulation are derived from the previous works[39].

4.2 Fabrication of the micrometer-scale nanophotonic film array

The fabrication procedure of the micrometer-scale nanophotonic film array starts with electron beam evaporation (Optorun, OFTC-1300) of the bottom three layers on a 4-inch silica substrate with a thickness of 500 μm. The intermediate layer is fabricated by a four-cycle process comprising photolithography (EV Group, EVG 620), magnetron sputtering (Kurt. J. Lesker, PVD 75), and stripping (acetone, aided by 400-watt ultrasonic agitation, is employed to effectively remove the photoresist and the overlying film). By varying the pattern of the photolithography masks and the deposition time, the nanophotonic film array consisting of 16 different cavity thicknesses is realized. Additionally, by altering the exposure dose for UV lithography, the morphology of the negative photoresist is manipulated to achieve a desired degree of trapezoidality in the resulting pattern. The top three layers are also fabricated using electron beam evaporation, and the whole configuration of the 7-layer ultracompact film stack comprises alternating high and low refractive index layers of $TiO_2$ and $SiO_2$, specifically structured as $TiO_2$ (60 nm) | $SiO_2$ (94 nm) | $TiO_2$ (60 nm) | $SiO_2$ (variable) | $TiO_2$ (60 nm) | $SiO_2$ (94 nm) | $TiO_2$ (60 nm).

4.3 Integration of the nanophotonic film array on the CIS

The cover glass of the monochromatic industrial camera (Hikrobot, MV-CB060-10UM-C, with Sony IMX178 sensor) is removed initially. The micrometer-scale nanophotonic film array is cut into 5 mm × 5.5 mm using mechanical cutting technique. Following this, the nanophotonic film array is bonded to the sensor using the optical adhesive (Norland, NOA61) with the aid of the transfer platform (Metatest, E1-G).

4.4 Characterization of the nanophotonic film array samples

Transmission spectra of the 4 × 4 nanophotonic film array are measured by the commercial spectrophotometer (CRAIC, 20/30PV), and the incident light spots are set as 7 μm × 7 μm for each individual film unit. Morphology of the film array is examined using Scanning Electron Microscopy (Zeiss, Ultra 55), and Atomic Force Microscopy (Bruker, Dimension ICON). The reflection spectra of the testing samples, e.g. the 24-color chart, the windmill, and the tangram, are measured by the commercial spectrometer (Hitachi, UH4150 with the integrating sphere) and the spectrophotometer (CRAIC, 20/30PV).

4.5 Calibration of the hyperspectral imager and the spectral imaging set up

The spectral responses of the all pixels are calibrated using a 400-watt Xenon lamp (Microenerg, CME-SL150) as a broadband light source and a monochromator (Zolix, Omni-λ5005i). The light is expanded and collimated before directed onto the CIS under normal incidence, ensuring that the light spot received by the CIS is uniform. The experiment setup for calibration is shown in Fig. S12, Supporting Information. During hyperspectral imaging experiments, the imager is equipped with an imaging lens of varying focal lengths (Hikrobot, MVL-KF2528M-12MPE and Hikrobot, MVL-HF0828M-6MPE) to accommodate different fields of view (FOVs) and object distances. In indoor imaging experiments, the aforementioned Xenon lamp is employed as the light source, and the experiment setup for indoor spectral imaging is shown in Fig. S13, Supporting Information. And the reference RGB images are captured using a RGB industrial camera (Hikrobot, MV-CB060-10UC-C, with Sony IMX178 sensor).

**References**
1. Lee, Z., Carder, K. L., Mobley, C. D., Steward, R. G. & Patch, J. S. Hyperspectral remote sensing for shallow waters. I. A semianalytical model. *Appl. Opt.* **37**, 6329–6338 (1998).
2. Dale, L. M. et al. Hyperspectral Imaging Applications in Agriculture and Agro-Food Product Quality and Safety Control: A Review. *Appl. Spectrosc. Rev.* **48**, 142-159 (2013).
3. Lebourgeois, V. et al. Can commercial digital cameras be used as multispectral sensors? A crop

monitoring test. *Sensors* **8**, 7300–7322 (2008).

4. Phang, S. K., Chiang, T. H. A., Happonen, A. & Chang, M. M. L. From Satellite to UAV-Based Remote Sensing: A Review on Precision Agriculture. *IEEE Access* **11**, 127057–127076 (2023).

5. Lu, G. & Fei, B. Medical hyperspectral imaging: a review. *J. Biomed. Opt.* **19**, 010901 (2014).

6. Liang, H. Advances in multispectral and hyperspectral imaging for archaeology and art conservation. *Appl. Phys. A* **106**, 309–323 (2012).

7. Shaw, G. A. & Burke, H. K. Spectral Imaging for Remote Sensing. *Lincoln Lab. J* **14**, 3–28 (2003).

8. Lin, X., Liu, Y., Wu, J. & Dai, Q. Spatial-spectral encoded compressive hyperspectral imaging. *ACM Trans. Graph.* **33**, 1-11 (2014).

9. Gehm, M. E., John, R., Brady, D. J., Willett, R. M. & Schulz, T. J. Single-shot compressive spectral imaging with a dual-disperser architecture. *Opt. Express* **15**, 14013–14027 (2007).

10. Golub, M. A. et al. Compressed sensing snapshot spectral imaging by a regular digital camera with an added optical diffuser. *Appl. Opt.* **55**, 432-443 (2016).

11. Peng, Y. F. et al. The diffractive achromat full spectrum computational imaging with diffractive optics. *ACM Trans. Graph.* **35**, 31 (2016).

12. Huang, L., Luo, R., Liu, X. & Hao, X. Spectral imaging with deep learning. *Light: Sci. Appl.* **11**, 61 (2022).

13. Yang, Z. et al. Single-nanowire spectrometers. *Science* **365**, 1017–1020 (2019).

14. Meng, J., Cadusch, J. J. & Crozier, K. B. Detector-Only Spectrometer Based on Structurally Colored Silicon Nanowires and a Reconstruction Algorithm. *Nano Lett.* **20**, 320–328 (2020).

15. Yuan, S., Naveh, D., Watanabe, K., Taniguchi, T. & Xia, F. A wavelength-scale black phosphorus spectrometer. *Nat. Photonics* **15**, 601–607 (2021).

16. Guo, L. et al. A Single-Dot Perovskite Spectrometer. *Adv. Mater.* **34**, 2200221 (2022).

17. Yoon, H. H. et al. Miniaturized spectrometers with a tunable van der Waals junction. *Science* **378**, 296–299 (2022).

18. Bao, J. & Bawendi, M. G. A colloidal quantum dot spectrometer. *Nature* **523**, 67–70 (2015).

19. Zhu, X. et al. Broadband perovskite quantum dot spectrometer beyond human visual resolution. *Light: Sci. Appl.* **9**, 73 (2020).

20. Craig, B., Shrestha, V. R., Meng, J., Cadusch, J. J. & Crozier, K. B. Experimental demonstration of infrared spectral reconstruction using plasmonic metasurfaces. *Opt. Lett.* **43**, 4481–4484 (2018).

21. Kim, C., Ni, P., Lee, K. R. & Lee, H.-N. Mass production-enabled computational spectrometers based on multilayer thin films. *Sci. Rep.* **12**, 4053 (2022).

22. Huang, E., Ma, Q. & Liu, Z. Etalon Array Reconstructive Spectrometry. *Sci. Rep.* **7**, 40693 (2017).

23. Wang, Z. et al. Single-shot on-chip spectral sensors based on photonic crystal slabs. *Nat. Commun.* **10**, 1 (2019).

24. Qu, Y., Zhou, Q., Xiang, J. & Yu, Z. Sparsity for Ultrafast Material Identification. Preprint at https://doi.org/10.48550/arXiv.2212.13122 (2022).

25. Zhu, Y., Lei, X., Wang, K. X. & Yu, Z. Compact CMOS spectral sensor for the visible spectrum. *Photonics Res.* **7**, 961–966 (2019).

26. Brown, C. et al. Neural Network-Based On-Chip Spectroscopy Using a Scalable Plasmonic Encoder. *ACS Nano* **15**, 6305–6315 (2021).

27. Xiong, J. et al. Dynamic brain spectrum acquired by a real-time ultraspectral imaging chip with reconfigurable metasurfaces. *Optica* **9**, 461 (2022).
28. Yang, J. et al. Ultraspectral Imaging Based on Metasurfaces with Freeform Shaped Meta-Atoms. *Laser Photonics Rev.* **16**, 2100663 (2022).
29. Rao, S., Huang, Y., Cui, K. & Li, Y. Anti-spoofing face recognition using a metasurface-based snapshot hyperspectral image sensor. *Optica* **9**, 1253 (2022).
30. Yang, J. et al. Deep-learning based on-chip rapid spectral imaging with high spatial resolution. *Chip* **2**, 100045 (2023).
31. Yako, M. et al. Video-rate hyperspectral camera based on a CMOS-compatible random array of Fabry–Pérot filters. *Nat. Photonics* **17**, 218–223 (2023).
32. Wen, J. et al. Deep Learning-Based Miniaturized All-Dielectric Ultracompact Film Spectrometer. *ACS Photonics* **10**, 225–233 (2023).
33. Zhang, W. et al. Deeply learned broadband encoding stochastic hyperspectral imaging. *Light: Sci. Appl.* **10**, 108 (2021).
34. Song, H. et al. Deep-Learned Broadband Encoding Stochastic Filters for Computational Spectroscopic Instruments. *Adv. Theory Simul.* **4**, 2000299 (2021).
35. Yasuma, F., Mitsunaga, T., Iso, D. & Nayar, S. K. Generalized Assorted Pixel Camera: Postcapture Control of Resolution, Dynamic Range, and Spectrum. *IEEE Trans. Image Proc.* **19**, 2241–2253 (2010).
36. Arad, B. & Ben-Shahar, O. Sparse Recovery of Hyperspectral Signal from Natural RGB Images. in Proceedings of the 14th European Conference on Computer Vision (eds Leibe, B., Matas, J., Sebe, N. & Welling, M.) 19–34 (Cham, 2016).
37. Yang, C. et al. Angle Robust Reflection/Transmission Plasmonic Filters Using Ultrathin Metal Patch Array. *Adv. Opt. Mater.* **4**, 1981–1986 (2016).
38. Fang, B. et al. Highly efficient omnidirectional structural color tuning method based on dielectric–metal–dielectric structure. *Appl. Opt.* **56**, C175–C180 (2017).
39. Wen, J. et al. Thin film-based colorful radiative cooler using diffuse reflection for color display. *PhotoniX* **4**, 25 (2023).